\documentclass[aps,prl,showpacs,amsmath,twocolumn]{revtex4}
\usepackage{graphicx}
\usepackage{wrapfig}
\usepackage{amsmath}

\def \nup {n_{\uparrow}}
\def \ndo {n_{\downarrow}}

\def\si{\sigma}
\def\oneh{\frac{1}{2}}

\newcommand{\intlim}{\mathop{\int}\limits}

\begin{document}
\title{The relationship between Hirsch-Fye and weak coupling
 diagrammatic Quantum Monte Carlo methods.}
\author{K. Mikelsons}
\email{mikelsk@email.uc.edu}
\affiliation{Department of Physics, University of Cincinnati,
 Cincinnati, Ohio 45221, USA}
\affiliation{Department of Physics and Astronomy, 
Louisiana State University, Baton Rouge, Louisiana 70803, USA}
\author{A. Macridin}
\affiliation{Department of Physics, University of Cincinnati,
 Cincinnati, Ohio 45221, USA}
\author{M. Jarrell}
\affiliation{Department of Physics and Astronomy, 
Louisiana State University, Baton Rouge, Louisiana 70803, USA}
\date{\today}

\begin{abstract}
Two weak coupling Continuous Time Quantum Monte Carlo (CTQMC) methods
 are shown to be equivalent for Hubbard-like interactions. A relation
 between these CTQMC methods and the Hirsch-Fye Quantum Monte Carlo
 (HFQMC) method is established, identifying the latter as an
 approximation within CTQMC and providing a diagrammatic interpretation.
 Both HFQMC and CTQMC are shown to be equivalent when the number of
 time slices in HFQMC becomes infinite, implying the same degree of
 fermion sign problem in this limit. 
\end{abstract}
\maketitle

{\em{Introduction.}} \label{sec:intro}
Hirsch-Fye Quantum Monte Carlo (HFQMC) is a standard method for the
 simulation of quantum lattice 
models~\cite{hirschfye86, fyehirsch88, georges96, assaad02, santos03}.
 However, during the past decade, new QMC methods have emerged, which
 are based on stochastic sampling of diagrams in a perturbative
 expansion~\cite{prokofev96}. These new methods avoid
 systematic errors due to finite discretization in the imaginary
 time, and are commonly referred to as "Continuous Time QMC"
 (CTQMC). With new variants of CTQMC appearing, a comparison of
 these formalisms becomes important. 
In this work, we consider two seemingly different CTQMC methods
 proposed by Rombouts~\cite{rombouts99}  and Rubtsov~\cite{rubtsov05}
 based on the expansion of the interaction term of Hamiltonian in the
 perturbation series, also known as "weak coupling" CTQMC.  We show
 that for Hubbard-like interactions these methods are equivalent.
 We also show that HFQMC can be interpreted as a summation of
 a specific subset of diagrams present in CTQMC.

{\em The equivalence of CTQMC algorithms
 by Rombouts and Rubtsov.}\label{sec:CTQMC_CTAUX}
These methods consider the perturbative expansion of the partition
 function in  powers of the interaction and then sample the resulting
 series of multi-dimensional integrals stochastically. 
We will use a path integral formalism to illustrate this. Here,
 the partition function is written as an integral over the Grassman
 variables $\eta, \eta^*$:
$Z = \int \mathcal{D} \eta^* \mathcal{D} \eta e^{-S(\eta^*,\eta)}$,
 with the action 
\begin{equation}
S(\eta^*,\eta) = S_0(\eta^*,\eta) -\int_0^{\beta}
 d\tau V(\eta^*(\tau),\eta(\tau)) \,,
\end{equation}
where $S_0$ is the bare part of $S$, and $V$ is the
interacting part of the Hamiltonian $H$.
For the purposes of discussion we consider Hubbard like repulsive
 interaction~\cite{hubbard}:
\begin{equation}
V = U \sum_{j=1}^{N_c} \left[ n_{j\uparrow} n_{j\downarrow} -
 \oneh(n_{j\uparrow} + n_{j\downarrow})\right] \,.
\end{equation} 

In the Rombouts method~\cite{rombouts99}, a constant $K$ is
 introduced to shift the reference free energy and the resulting
 series expansion for the partition function can be written as:
\begin{eqnarray}
Z  &=& e^{-K} \int_{\eta^*,\eta}  e^{-S_0} \sum_{k=0}^\infty
 \left(\frac{K}{\beta}\right)^k \int_0^{\beta} d\tau_1 \ldots
 \int_0^{\tau_{k-1}} d\tau_k \times \nonumber \\
 &&  \times \left( 1 - \frac{\beta}{K} V(\tau_1) \right)
 \ldots \left( 1 - \frac{\beta}{K} V(\tau_k) \right) \,.
\end{eqnarray}
The following identity is then used to decouple the interaction
 terms and introduce an auxiliary field $s$:
\begin{equation}
\left( 1 - \frac{\beta}{K} V \right) =  \frac{1}{2 N_c}
 \sum_{j=1}^{N_c} \sum_{s_j=\pm1}
  e^{\gamma s_j (n_{j\uparrow} - n_{j\downarrow})}
\label{eq:RHHS}  \,,
\end{equation}
where $\cosh\gamma = 1 + \frac{\beta U N_c}{2 K}$.
 The resulting series for the partition function is:
\begin{equation}
\begin{split}
Z &= e^{-K} \int_{\eta^*,\eta}  e^{-S_0}
 \sum_{\substack{k j  \tau s}} 
 \left(\frac{K}{2 \beta N_c}\right)^k  \times \\
 & \times e^{\gamma s_1 [n_{j_1 \uparrow}
 (\tau_1) - n_{j_1 \downarrow} (\tau_1)]} 
 \ldots e^{\gamma s_k [n_{j_k \uparrow}
 (\tau_k) - n_{j_k \downarrow}(\tau_k)] } \,,
\label{eq:Z_Rombouts}
\end{split}
\end{equation}
where multiple sums and integrals are denoted as:
\begin{equation}
\sum_{\substack{k j \tau s}} = \sum_{k=0}^\infty
 \int_0^{\beta} d\tau_1 \sum_{j_1=1}^{N_c} \sum_{s_1} \ldots 
\int_0^{\tau_{k-1}} d\tau_k \sum_{j_k=1}^{N_c} \sum_{s_k} \,.
\label{eq:multisum}
\end{equation}
The fermion degrees of freedom can now be integrated out, and
 the partition function can be rewritten as~\cite{gull08}:
\begin{equation}
Z = \frac{Z_0}{e^{K}} \sum_{\substack{k j  \tau s}}
  \left(\frac{K}{2 \beta N_c}\right)^k  \prod_{\si}
 \det {\mathcal{G}^0_{\si}} \cdot \det \left[ G_{\si}^{\{s_i\}}
 \right]^{-1} \,,
\label{eq:RZ}
\end{equation}
where $G_{\si}^{\{s_i\}}$ is the Green's function for a particular
 configuration of auxiliary fields, and is related to the
 non-interacting Green's function $\mathcal{G}^0_{\si}$ by
 a Dyson's equation:
\begin{equation}
\left[ G_{\si}^{\{s_i\}} \right]^{-1} 
 e^{-\gamma W_{\si}^{\{s_i\}}} - e^{-\gamma W_{\si}^{\{s_i\}}} =
 \left[\mathcal{G}^0_{\si}\right]^{-1} - \mathrm{I} 
\label{eq:Rombouts_update}
\end{equation}
with $W_{\si}^{\{s_i\}} = \mathrm{diag} ( \si s_i)$ and
 $\left[{\mathcal{G}^0_{\si}}\right]_{pq} =
 \mathcal{G}^0_{\si}(j_p, \tau_p; j_q, \tau_q)$ being
 $k \times k$ matrices. 
Finally, QMC is used to perform the multidimensional sum
 (Eq.~\ref{eq:multisum}) over different expansion orders
 and configurations of the auxiliary fields. For this, a
 Markov process is set up that samples the configurations
 of random auxiliary fields $\{s_i\}$ with weight given by
 the product of determinants in Eq.~\ref{eq:RZ}. 

In the Rubtsov method~\cite{rubtsov05, assaad07},
 the interaction is first rewritten as:
\begin{equation}
V' = \frac{U}{2} \sum_{j=1}^{N_c} \sum_{\tilde{s}_j=\pm1}
 \left( n_{j\uparrow} - \oneh -\alpha \tilde{s}_j\right)
 \left( n_{j\downarrow} - \oneh + \alpha \tilde{s}_j\right) \,,
\label{eq:Rubtsov_V}
\end{equation}
thereby introducing auxiliary fields $\tilde{s}$. This
 amounts to introducing a shift in the free energy:
\begin{equation}
K = \beta U N_c \left(\alpha^2-\frac{1}{4}\right) \,.
\label{eq:RRequiv_mu_alpha}
\end{equation} 
The auxiliary fields $\tilde{s}$ suppress the oscillating
 sign of the integrand in the perturbative
 expansion~\cite{rubtsov05}:
\begin{equation}
\begin{split}
\label{eq:Z_Rubtsov}
&Z  = e^{-K} \int_{\eta^*,\eta} e^{-S_0}
  \sum_{\substack{k j \tau s}}  \left(\frac{-U}{2}\right)^k
  \prod_{\si}  \\ 
& \left[ n_{j_1 \si}(\tau_1) - \oneh - \alpha \si
 \tilde{s}_1 \right] \ldots \left[ n_{j_k \si}(\tau_k)
 - \oneh - \alpha \si \tilde{s}_k \right]  
\end{split} 
\end{equation}
The fermion degrees of freedom can be integrated out, and
 the partition function becomes~\cite{assaad07}:
\begin{equation}
Z =  \frac{Z_0}{e^{K}} \sum_{\substack{k j  \tau s}}
 \left(\frac{-U}{2}\right)^k \prod_{\si} \det
 \left({\mathcal{G}^0_{\si}} - \oneh -
 \alpha W_{\si}^{\{\tilde{s}_i\}} \right)  
\label{eq:Z_Rubtsov_det}
\end{equation} 
Again, the product of determinants in
 Eq.~\ref{eq:Z_Rubtsov_det} gives the weight in QMC to evaluate 
the multidimensional sum over the configurations of auxiliary
 fields $\{\tilde{s}_i\}$. 

We now show that the two expansions (\ref{eq:RZ}) and
 (\ref{eq:Z_Rubtsov_det}) are equivalent (term by term) and
 that the auxiliary fields $\{s_i\}$ and $\{\tilde{s}_i\}$ are
 equivalent as well. Using Eq.~\ref{eq:Rombouts_update}, the
 inverse Green's function ${G^{\{s_i\}}_{\si}}^{-1}$ can be
 rewritten as:
\begin{equation}
{\mathrm{G}_{\si}^{\{s_i\}}}^{-1} =  {\mathcal{G}^0_{\sigma}}^{-1}
 \left[{\mathcal{G}^0_{\sigma}} - \frac{\mathrm{I}}{2} - 
 \alpha^* \mathrm{W}_{\si}^{\{s_i\}} \right] 
\left( \mathrm{I}-e^{\gamma \mathrm{W}_{\si}^{\{s_i\}} } \right) \,,
\end{equation}
where $\alpha^* = \left[2\tanh \frac{\gamma} {2}\right]^{-1} $.
Using this, and the fact that $\prod_{\si} (1-e^{\gamma \si s_i})
 = 2-2\cosh \gamma = -\frac{\beta U N_c}{K}$, the integrand of the
 Eq.~\ref{eq:RZ}, can be rewritten as: 
\begin{equation}
\begin{split}
&\left(\frac{K}{2\beta N_c}\right)^k  \prod_{\si}
 \det {\mathcal{G}^0_{\si}} \cdot
 \det \left[G_{\si}^{\{s_i\}}\right]^{-1} = \\ 
& = \left(-\frac{U}{2}\right)^k \prod_{\si}
 \det \left( {\mathcal{G}^0_{\si}} - \oneh
 - \alpha^* W_{\si}^{\{s_i\}} \right) \,,
\end{split}  
\end{equation} 
from which we deduce that both algorithms are equivalent if
 $\alpha = \alpha^*$, which is the same as requiring that
 Eq.~\ref{eq:RRequiv_mu_alpha} holds for freely adjustable
 parameters $K$ and $\alpha$ in these methods. 
Both algorithms must have the same degree of sign problem
 and statistics of measurements (such as auto-correlation time),
 as long as the above mentioned condition for the parameters
 $K$ and $\alpha$ is satisfied. 

\begin{figure}[h]
\begin{center}
\includegraphics[width=8.0cm]{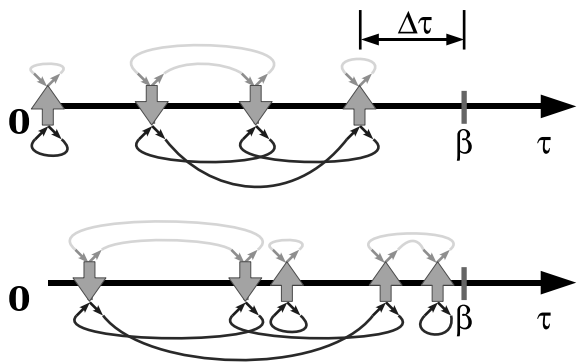}
\caption{Example of the diagrams sampled in HFQMC (top) and
 CTQMC (bottom). The light and dark lines denote propagation
 of spin-up and spin-down fermions, respectively. The block
 arrows represent the auxiliary field, associated with an
 interaction vertex. These vertices are distributed evenly
 in HFQMC, with interval $\Delta \tau$, while the distribution,
 positions and the number of vertices are arbitrary in the CTQMC.
 Note that both connected and disconnected diagrams are sampled
 in both methods. In fact, product of determinants
 (Eqs.~\ref{eq:Z_Rubtsov_det} and~\ref{eq:PATHINT_Zdet}) accounts
 for all $(k!)^2$ possible diagrams for a particular set of
 vertices.}
\label{fig:HFQMC_CTQMC_diagrams}
\end{center}
\end{figure}

{\em The relation between HFQMC and CTQMC.}
\label{sec:HFQMC_CTQMC}
The derivation of the Hirsch-Fye algorithm involves breaking
 up the partition function using a Trotter decomposition and
 decoupling the quartic part of the Hamiltonian with the
 transformation~\cite{hirsch83,fyehirsch88}:
\begin{equation}
e^{-\Delta \tau U \left[\nup\ndo - \oneh(\nup+\ndo) \right]} =
 \oneh \sum_{s=\pm1} e^{\lambda s (\nup-\ndo)} \,,
\label{eq:HFQMC_HHS_transf}
\end{equation}
where $\cosh\lambda = e^{\Delta \tau U /2}$. The resulting
 partition function takes the
 well-known form~\cite{hirschfye86, fyehirsch88}:
\begin{eqnarray}
Z &=& \sum_{\{s_j\}} \int_{\eta^*,\eta}
 e^{-\sum_{\substack{ij\si}} \eta_{i\si}^*
 \left[\left( {\mathcal{G}^0_{\si}}^{-1} -\mathrm{I} \right)
  e^{\lambda W_{\si}^{\{s_j\}}} +
 \mathrm{I} \right] \eta_{j\si} } \\
  &=& \sum_{\{s_j\}} \prod_{\sigma=\pm1}
 \det \left[ G_{\si}^{\{s_j\}}\right]^{-1} \ , 
\label{eq:PATHINT_Zdet}
\end{eqnarray}
where $G_{\si}^{\{s_j\}}$ is the Green's function for a
 particular configuration of auxiliary fields $\{s_j\}$:
\begin{equation}
\left[ G_{\si}^{\{s_j\}} \right]^{-1} =
 {\mathcal{G}^0_{\sigma}}^{-1} e^{\lambda W_{\si}^{\{s_j\}}}
 - e^{\lambda W_{\si}^{\{s_j\}}}  +  \mathrm{I} \,.
\label{eq:PATHINT_G_sj} 
\end{equation}
The product of determinants (\ref{eq:PATHINT_Zdet}) yields
 a sampling weight for a corresponding configuration of the
 auxiliary fields $\{s_j\}$. 
It is very similar to the sampling weight in CTQMC
 (Eq.~\ref{eq:RZ}), as are the transformations employed
 (Eqs.~\ref{eq:HFQMC_HHS_transf} and \ref{eq:RHHS}) and the
 update formulas (Eq.~\ref{eq:PATHINT_G_sj} and
 \ref{eq:Rombouts_update}). The only difference is that in
 HFQMC, the number of the auxiliary fields is fixed to
 $k_{\textrm{HF}} = \beta N_c / \Delta \tau$, and they are
 distributed evenly in the imaginary time. In addition, the
 parameter $\lambda$ plays the same role as the parameter
 $\gamma$ in CTQMC to couple the auxiliary fields to the
 fermion spin. In fact, one can formulate a set of restrictions,
 under which CTQMC reduces to HFQMC:
\begin{enumerate}
	\item Restrict the expansion order $k$ in CTQMC equal to
 the number of the auxiliary fields $k_{\textrm{HF}}$ in the
 HFQMC and distribute them evenly in the imaginary time
 interval $(0\ldots\beta)$.
	\item Set the strength of the auxiliary field in CTQMC:
 $\gamma = \lambda$. In terms of CTQMC parameters $K$ or
 $\alpha$, this condition is equivalent to:
	\begin{eqnarray}
         K &=& \frac{\beta U N_c}{2 \sinh^2 \frac{\lambda}{2}} =
 \frac{\beta U N_c}{2\left(e^{\frac{\Delta\tau U}{2}}-1\right)} \,;
 \label{eq:K_of_lambda}\\
	\label{eq:lapha_k}
	\alpha &=& \frac{1}{2 \tanh \frac{\lambda}{2}} =
 \frac{1}{2\sqrt{\tanh \frac{\Delta \tau U}{4}}} \,.
 \label{eq:alpha_of_lambda}
	\end{eqnarray}
	\item Restrict the Monte-Carlo moves to flipping the
 auxiliary fields associated with the interaction vertices;
 shifting vertices in imaginary time is not allowed.
\end{enumerate}
These restrictions imply that only a subset of diagrams with
 fixed expansion order and equidistant auxiliary fields are
 sampled in HFQMC, whereas in CTQMC, all diagrams of variable
 order and all possible sets of auxiliary field configurations
 contribute (see Fig.~\ref{fig:HFQMC_CTQMC_diagrams}).

The relation between CTQMC and HFQMC persists for the attractive
 Hubbard model ($U<0$). Here, a different discrete
 Hubbard Stratonovich transformation is used:
\begin{equation}
e^{\Delta \tau |U| \left[\nup\ndo - \oneh(\nup+\ndo-1) \right]}
 = \oneh \sum_{s=\pm1} e^{\lambda s (\nup + \ndo - 1)} \,
\label{eq:HFQMC_HHS_transf_negU}
\end{equation}
with $\cosh \lambda=e^{\Delta \tau |U| /2}$. The corresponding
 form for the interaction in CTQMC also has to be modified:
\begin{equation}
V' = \frac{U}{2} \sum_{j=1}^{N_c} \sum_{\tilde{s}_j=\pm1}
 \left( n_{j\uparrow} - \oneh + \alpha \tilde{s}_j\right)
 \left( n_{j\downarrow} - \oneh + \alpha \tilde{s}_j\right)  \,.
\label{eq:Rubtsov_V_attr}
\end{equation}
Since the attractive Hubbard model has no sign problem, the
 parameter $\alpha$ can be set equal to zero. However, for
 $\alpha>0$, the relation to HFQMC is again given by the same
 set of restrictions as defined above (including
 Eqs.~\ref{eq:K_of_lambda},\ref{eq:alpha_of_lambda}). Similarly,
 the relation between CTQMC and HFQMC is preserved in case of
 nonlocal density-density interactions.

{\em Small $\Delta \tau$ limit.} \label{sec:dtau0_limit}
When $\Delta \tau \rightarrow 0$, systematic errors in HFQMC
 are eliminated and in this sense HFQMC and CTQMC are equivalent.
 The relationship described above will also hold for
 $\alpha \rightarrow \infty$ (see Eq.~\ref{eq:alpha_of_lambda}).
In the discussion above, HFQMC is interpreted as sampling just
 one order in series expansion. To understand this, we need to
 revisit the sampling and measurement procedure in the CTQMC.
 The expectation value of any operator can be written as a
 series expansion:
\begin{equation}
G = \frac{1}{Z}\sum_k G_k = \frac{1}{Z}\sum_k \frac{G_k}{Z_k}
 Z_k = \frac{1}{Z}\sum_k g_k Z_k \,.
\end{equation}
In both variants of CTQMC the evaluation of this sum is done
 with importance sampling, and the weight (or the "guiding
 function") is taken to be equal to the corresponding contribution
 to the partition function $Z_k$, with $g_k=\frac{G_k}{Z_k}$ being
 the Monte Carlo estimator for a fixed order of expansion. 
Of course, $Z_k$ depends on the configuration of the auxiliary
 fields $\{s_k\}$, so the actual estimator is $g_k^{\{s_k\}} =
 \frac{G_k^{\{s_k\}}}{Z_k^{\{s_k\}}}$. However, for this
 discussion, we are only interested in how this estimator depends
 on the expansion order, so we assume that the auxiliary fields
 are already summed.

\begin{figure}[h]
\begin{center}
\includegraphics[width=8.5cm]{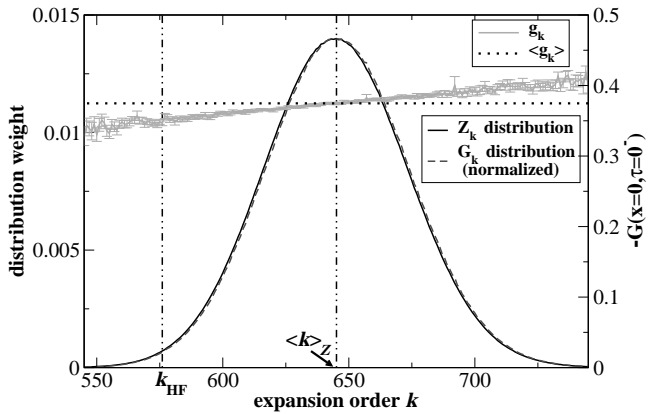} 
\caption{The distribution of the contributions $Z_k$ ($G_k$) to
 the partition function (Green's function) vs. the expansion
 order $k$ for the 2D Hubbard model with $N_c=16$ sites, $U=W$
 (bandwidth), filling $f=0.75$, $\beta=4.5$, $\alpha=1.01$.
 Note that the distributions almost perfectly overlap, and their
 ratio (proportional to $g_k$) varies very little with $k$.
 Restricting diagrams to one order $k=k_{\mathrm{HF}}$ and
 restricting vertices to discrete imaginary time grid results
 in the Hirsh-Fye algorithm with $\Delta\tau t= 0.125$.}
\label{fig:k_distr}
\end{center}
\end{figure}

The series expansion for the partition function
 (Eq.~\ref{eq:Z_Rubtsov}) defines a distribution
 (see Fig.~\ref{fig:k_distr}) with mean value~\cite{assaad07}:
\begin{equation}
\left< k \right>_Z =-\int_0^{\beta} d\tau \left< V(\tau) \right> \,.
\end{equation}
This can be generalized for higher factorial moments:
\begin{equation}
\begin{split}
\left< (k)_n \right>_Z &= \left< k (k-1) \ldots (k-n+1) \right>_Z \\
= (-1)^n & \intlim_0^{\beta} d\tau_1 \ldots \intlim_0^{\beta}
 d\tau_n \left<T_{\tau} V(\tau_1) \ldots V(\tau_n)\right> \,.
\end{split}
\end{equation}
For the Hubbard model with sufficiently large $\alpha$,
 these moments scale as:
\begin{equation}
\lim_{\alpha \rightarrow \infty } \left< (k)_n \right>_Z =
 \left( \beta U N_c \alpha^2 \right)^n = \rho^n \,,
\label{eq:fact_moments_scaling}
\end{equation}
which is a property of Poisson distribution $P_{\rho}(k) =
 \frac{\rho^k e^{-\rho}}{k!}$ with parameter
 $\rho=\beta U N_c \alpha^2$. Of course, for large $\rho$, the
 Poisson distribution approximates a normal distribution
 (see Fig.~\ref{fig:k_distr}).
In a similar way as the series expansion for the partition
 function defines its distribution, the expansion for the
 Green's function (or any measurable quantity) defines
 another distribution:
\begin{eqnarray}
G(\tau_i,\tau_j) &=& \sum_k G_k(\tau_i,\tau_j) =
 \sum_k \frac{(-1)^k}{k!} \int_0^{\beta} d\tau_1 \ldots
  \nonumber \\
 \ldots \int_0^{\beta} d\tau_k &\cdot & \left< T_{\tau} c(\tau_i)
 c^{\dag} (\tau_j) V(\tau_1) \ldots V(\tau_k) \right>\,,
\end{eqnarray}
which is characterized by its factorial moments:
\begin{eqnarray}
\left< (k)_n \right>_{G_{ij}} &=& \frac{(-1)^n}{G(\tau_i,\tau_j)}
 \int_0^{\beta} d\tau_1 \ldots \int_0^{\beta} d\tau_n \times
 \nonumber \\ 
   && \times \left<T_{\tau} c(\tau_i) c^{\dag}(\tau_j) V(\tau_1)
 \ldots V(\tau_n)\right> \,.
\end{eqnarray}
In general, this distribution is different from the one defined
 by the expansion of the partition function.
However, in the limit when  $\alpha \rightarrow \infty$, the
 factorial moments of Green's function distribution scale as:
\begin{equation}
\lim_{\alpha \rightarrow \infty } \left< (k)_n \right>_G =
 \left( \beta U N_c \alpha^2 \right)^n =\lim_{\alpha
 \rightarrow \infty } \left< (k)_n \right>_Z \,.
\label{eq:fact_moments_scaling_G}
\end{equation}
Since all the moments for both distributions are the same,
 the distributions are the same as well in this limit, and
 the estimator $g_k$ becomes a constant, independent of the
 expansion order $k$ (see Fig.~\ref{fig:k_distr}). Thus, the
 sum over all expansion orders $k$ can be replaced by any single
 term corresponding to a fixed value of $k=k_{\mathrm{HF}}$.
 That explains why sampling just one single order in the
 expansion for the partition function (as is done in HFQMC)
 gives the same exact result when $\Delta\tau\rightarrow 0$.

\begin{figure}
\begin{center}
\includegraphics[width=7.5cm]{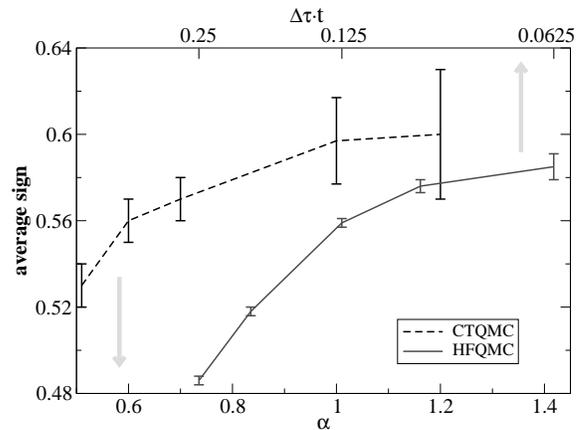}
\caption{Comparison of the average sign for the 2D Hubbard model
 with $N_c=24E$, $U=8t$, filling$=0.95$, $\beta t=4.5$. In CTQMC,
 the average sign depends on parameter $\alpha$ (lower horizontal
 axis), while in HFQMC, it depends on the size of the time slice
 $\Delta \tau t$ (upper horizontal axis). Marks on both axes are
 related by Eq.~\ref{eq:alpha_of_lambda}.
}
\label{fig:sign_CT_HF}
\end{center}
\end{figure}

{\em Computational implications.} \label{sec:sign}
When the product of determinants
 (Eqs.~\ref{eq:RZ},\ref{eq:Z_Rubtsov_det},\ref{eq:PATHINT_Zdet})
 is not positive definite, its absolute value is taken as a
 weight in QMC. This approach fails, if the average sign of the
 product of determinants becomes small. This is the infamous
 fermion sign problem, the main limitation in any fermion
 QMC method.
From the discussion above, it follows that both HFQMC and CTQMC
 have the same degree of sign problem when
 $\alpha\rightarrow \infty$ and $\Delta\tau\rightarrow 0$. For
 typical finite values of $\Delta \tau$ and $\alpha$, the
 difference in average sign is still small
 (see Fig.~\ref{fig:sign_CT_HF}) and depends on model parameters. 
Altogether, we find that neither of the methods has a definite
 advantage in terms of the degree of the sign problem.
Also, the auxiliary fields enter the same way in both methods,
 and correlations in these fields give information about the
 spin and charge  correlations in the repulsive and attractive
 Hubbard models, respectively. Thus, optimization strategies
 developed for HFQMC can be applied to CTQMC.

{\em Conclusions.} \label{sec:conclude}
We have investigated two weak coupling CTQMC methods proposed
 by Rombouts and Rubtsov, and shown that they are equivalent
 for a certain choice of freely adjustable parameters in these
 methods. We also established the relation between the CTQMC
 methods and HFQMC method and identified the latter as an
 approximation within CTQMC where the Monte Carlo sum is
 restricted to a certain subset of diagrams. We have shown
 that this approximation becomes exact in the limit when an
 infinite number of time slices is taken in HFQMC, implying
 that both methods have the same degree of the sign problem
 in this limit. 

{\em Acknowledgments.} \label{sec:acknowledgments}
We thank Emanuel Gull for useful discussions. The work was
 supported by NSF Grants Nos. DMR-0706379 and DMR-0312680.
 A.M. acknowledges the DOE CMSN Grant DE-FG02-04ER46129.


\end{document}